\documentclass[runningheads]{llncs}
\usepackage{graphicx}
\begin{document}
\title{A System Modeling Approach to Enhance Functional and Software Development}
\titlerunning{A System Modeling Approach to Enhance Functional and Software Development}
%
\author{Saurabh Tiwari\inst{1} \and Emina Smajlovic\inst{1} \and
Amina Krekic\inst{1} \and Jagadish Suryadevara\inst{2}}
\authorrunning{Tiwari et al.}
%
\institute{M\"{a}lardalen University, Sweden \and
Volvo Construction Equipment AB, Sweden \\
\email{saurabh.tiwari@mdh.se, \{esc17001, akc17003\}@student.mdh.se, jagadish.suryadevara@volvo.com}\\
}

\maketitle              

\begin{abstract}
This paper presents a SysML-based approach to enhance functional and software development process within an industrial context. The recent changes in technology such as electromobility and increased automation in heavy construction machinery lead to increased complexity for embedded software. Hence there emerges a need for new development methodologies to address flexible functional development, enhance communication among development teams, and maintain traceability from design concepts to software artifacts. The discussed approach has experimented in the context of developing a new transmission system (partially electrified) and its functionality. While the modeling approach is a \textit{work-in-progress}, some initial success, as well as existing gaps pointing to future works are highlighted.


\end{abstract}

\section{Introduction} 
\vspace{-0.7ex}
In recent times, there has been a significant paradigm shift within construction equipment industry in terms of introducing new technologies such as electromobility and increased automation. For instance, electrification (i.e. battery-powered parts) is being introduced into products, hydraulic motors are being replaced with electric versions, new versions of drive-line system (DLS) where electrified hub motors (instead of torque power from the engine) are introduced into wheels.

The advanced technological changes in large complex products causes enormous challenges for existing software development teams. While the functionality remains largely unchanged in comparison with legacy systems and software, the new \textit{design concepts} lead to major changes in hardware and software. Hence the traditional function development techniques largely based on small incremental changes to existing software is no longer valid and may lead to quality issues as well as maintainability and traceability problems. Model-based methodologies such as \textit{Model-based Systems Engineering} (\texttt{MBSE}) and \textit{Model-based Design} (\texttt{MBD}) are industry-wide considered as effective solutions in addressing, above described development challenges~\cite{friedenthal2007incose}\cite{friedenthal2014practical}. While traditionally MBD methodologies are associated with only Simulink-based development techniques, currently these techniques are being extended using SysML/UML-based modeling approaches. In this paper, we present the SysML-based modeling approach considered within the project for development of a new transmission system (partially electrified) and its software.

In this paper, we describe a modeling approach developed within VCE (Volvo
CE\footnote{Volvo Construction Equipment AB, Sweden}). The approach is presented using the \textit{`Brake'} functionality to capture the transmission system behavior, in response to \textit{brake requests} from the operator/machine. The modeling approach captures the functional behavior from both problem (as seen externally) and solution domain (albeit implementation independent) perspectives. Later, the SysML-based solution models can be further refined into \textit{hardware} and \textit{software views} explicitly reflecting the overall design concept(s), paving the way for traceable software architecture(s) which in turn implemented using traditional \textit{Simulink-based} techniques.

\section{Background}
Software Engineering Framework at VCE referred to as SE-Tool, is a customization of the commercial tool \textit{Systemweaver}. It is influenced by EAST-ADL\footnote{We refer the reader to the standard for further details: http://www.east-adl.info/} framework, to support complete software development processes at VCE. It is primarily used to develop the complete Electrical and Electronic (E2E) System in software as well as hardware i.e., ECUs, Sensors, and Actuators.

MathWorks Simulink\footnote{https://se.mathworks.com/products/simulink.html} is a graphical development tool used to run simulations, generate code, and test and verify embedded systems. All functionalities in the Simulink represented by the blocks. The Simulink connects different blocks and signals to simulation models that can be executed. Also, the block in a Simulink work similarly to functions in a C/C++ program. Blocks are divided into pre-defined MATLAB libraries\footnote{https://se.mathworks.com/help/simulink/block-libraries.html} based on their functionality. These libraries include Logic and Bit Operations (e.g., blocks like Relational Operator and Logical Operator) and Math Operations (e.g., blocks like Product, Divide, Add, Subtract are included).

While both SE-Tool and Simulink-based frameworks satisfactorily cover the
development processes within Software domain, currently there are huge gaps
related to systems engineering domain like maintainability, traceability, incorporating new design concepts etc. Hence, there is a need for the ``\textit{Model-based System}'' approach compared to traditional function development approach.

\section{Braking Functionality: An Example}
The drive-line system consists of a \textit{clutch}, a \textit{transmission}, a \textit{drive shaft}, and an \textit{axle} connecting the engine and the drive wheels. To describe the system modeling approach, we model the `Brake' functionality, based on the logical distribution of the system behavior in terms of the system components described above. Please note, to simplify the presentation of this paper both scope and all the figures and illustrations are limited to the behavior of the `Brake' functionality.

Corresponding to required ``Brake behavior'', the DLS system ``implements''
corresponding ``Braking'' behavior within its scope, e.g., to create a tractive force on the machine and gives the machine the capability to move. Additional sub-systems and components (besides described below) may be required to increase the quality of the movement, but they are skipped in this paper, to keep the presentation simple. In order to perform braking, mainly following subsystems and external entities are involved:

\begin{itemize}
    \item \textbf{Operator} represents a person who operates a machine by sending a ``braking request'' through appropriate interface. 
    \item \textbf{Powertrain} is a subsystem that includes following subsystems: Engine, Drivetrain and Wheels(\#4)
    \item \textbf{Requester} represents the braking pedal of the physical machine.
\end{itemize}

\section{SysML-based Modeling Approach}
\begin{figure*}[!ht]
\centerline{\includegraphics[height=1.78in,width=4.84in,angle=0]{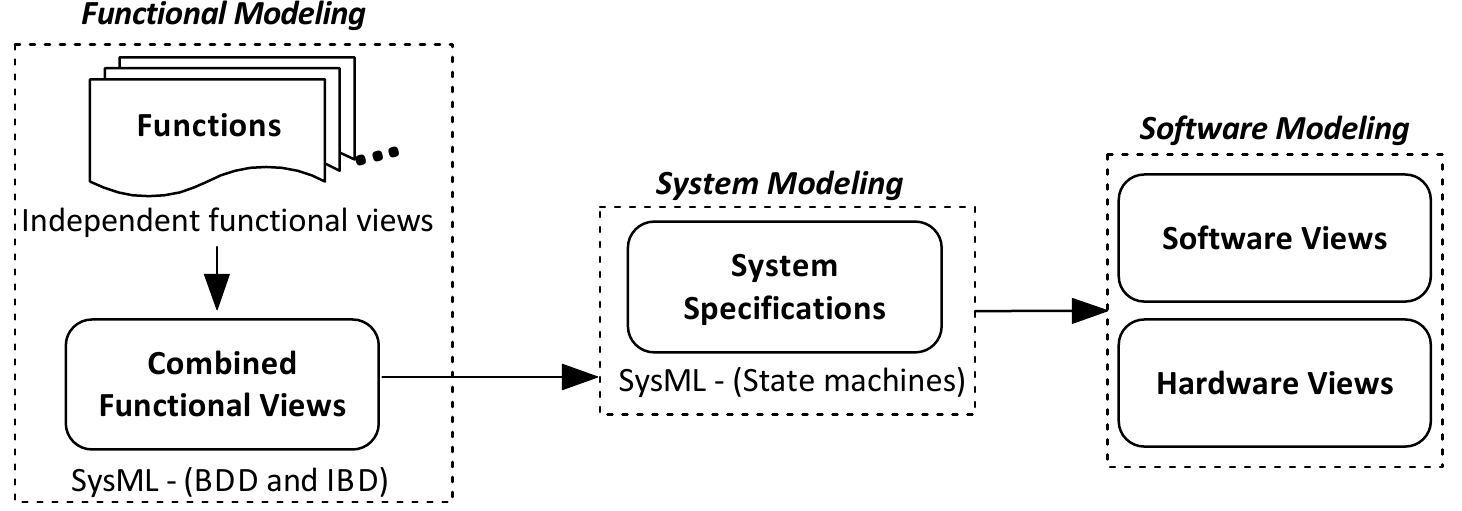}}
\vspace*{-1.6ex}
\caption{Overview of the SysML-based modeling approach}
\vspace*{-2.0ex}
\label{fig:approach}
\end{figure*}

The overall modeling methodology, as shown in Fig.~\ref{fig:approach}, is primarily based on a reverse-engineering approach, i.e., capturing functionality from legacy concepts and corresponding implementations in the hardware and software (e.g., simulink-
based environment). The modeling has been done using IBM Rhapsody tool\footnote{https://www.ibm.com/se-en/marketplace/rational-rhapsody}.
However, in this paper, for illustration purpose, the models have been re-drawn
using other tool.

The overall system modeling approach is divided into three stages.

\begin{figure*}[!ht]
\centerline{\includegraphics[height=2.5in,width=4.3in,angle=0]{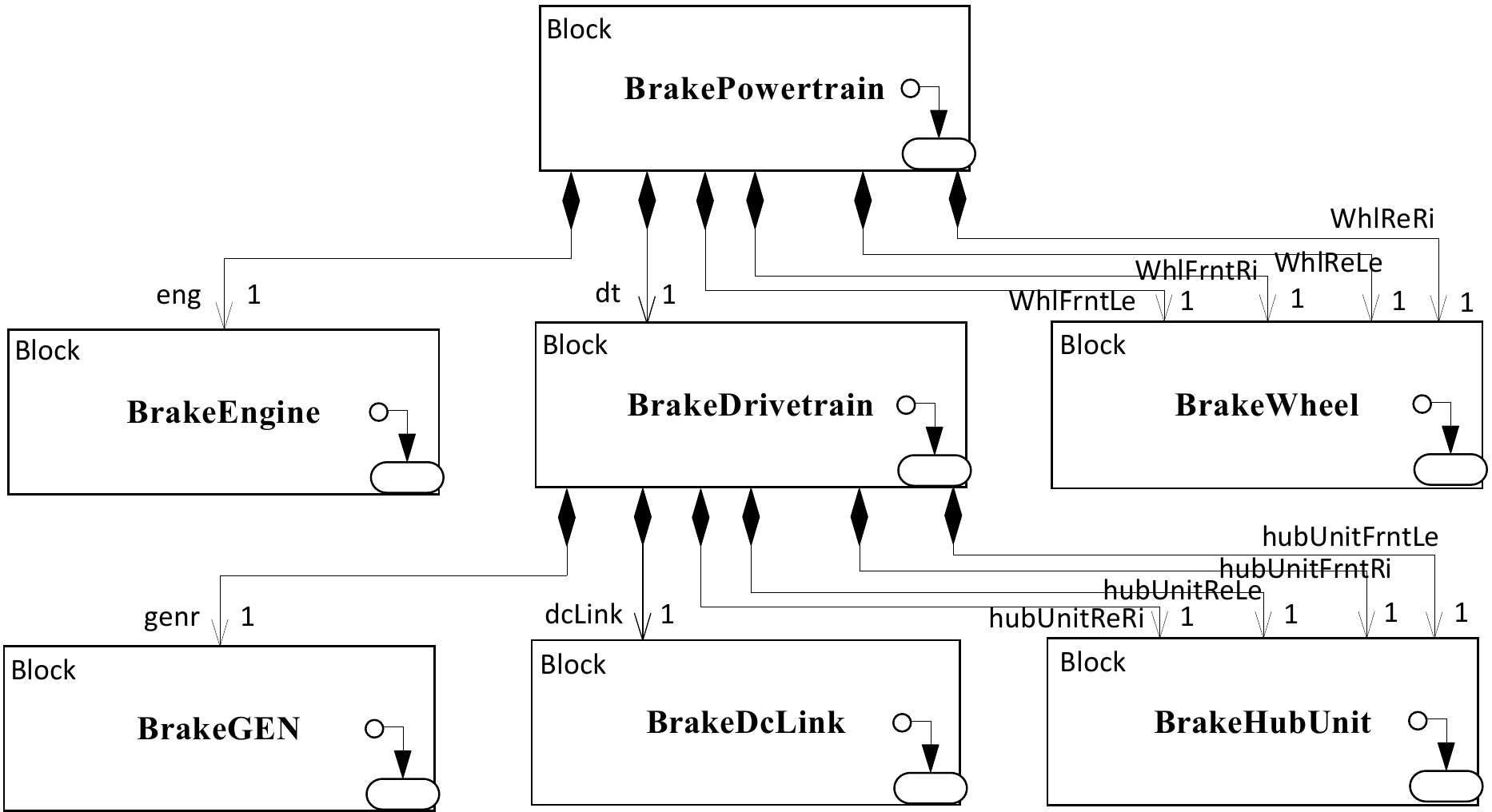}}
\vspace*{-1.6ex}
\caption{BDD for Powertrain subsystem: description of the \textit{Brake} function}
\vspace*{-2.0ex}
\label{fig:bdd}
\end{figure*}

\subsection{Functional Modeling (Structure)} 
\begin{figure*}[!ht]
\centerline{\includegraphics[height=2.77in,width=4.3in,angle=0]{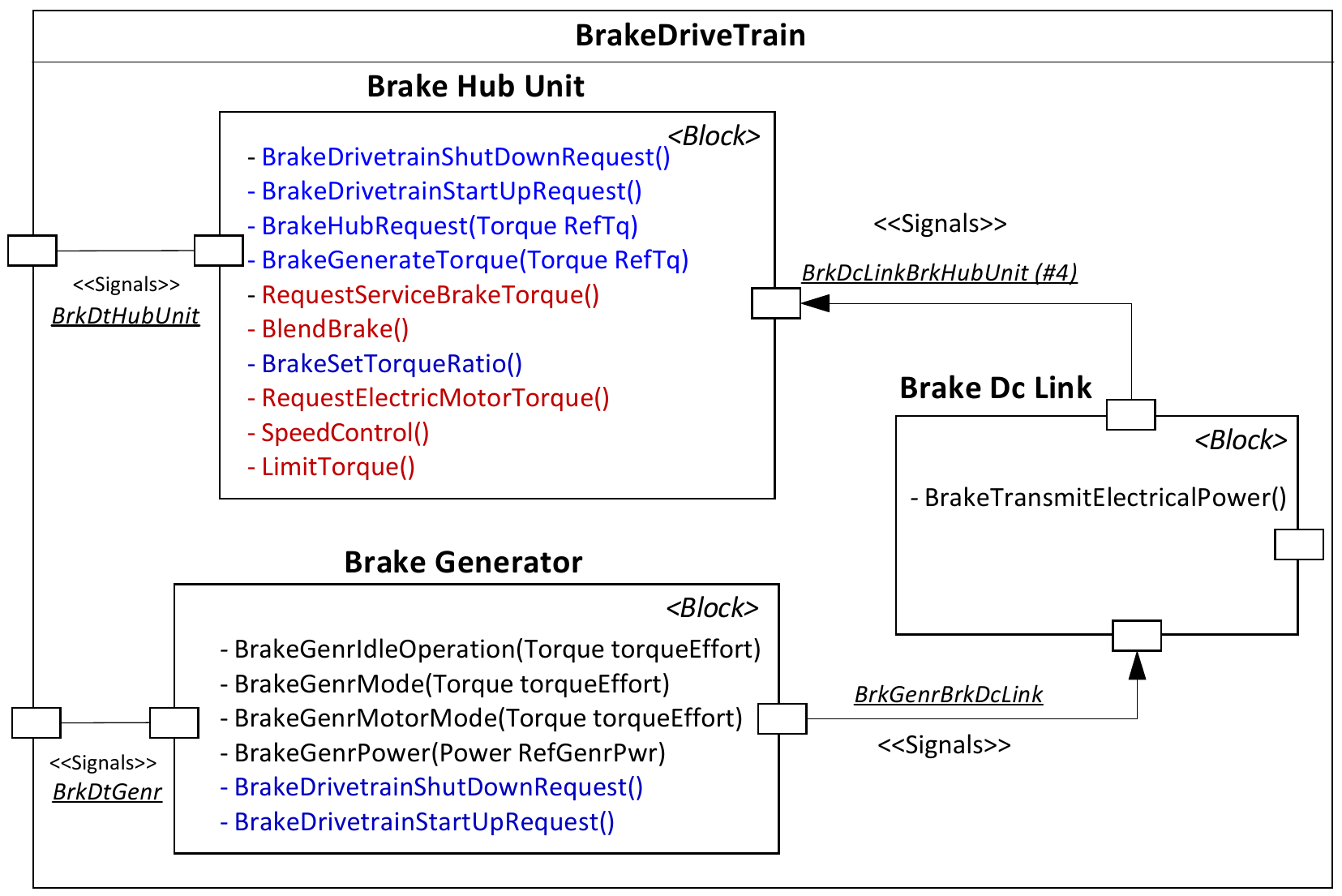}}
\vspace*{-2.0ex}
\caption{BrakeDrivetrain IBD representing \textit{Operations} and \textit{Signals}}
\vspace*{-3.4ex}
\label{fig:drivetrain}
\end{figure*}

The first phase of the modeling activity (refer to Fig.~\ref{fig:approach}), is based on the product breakdown structure. Thus the modeling focuses on ``structural aspects'' using SysML BDD (Block Definition Diagram) and IBD (Internal Block Definition) Diagrams (as illustrated in Fig.~\ref{fig:bdd} and Fig.~\ref{fig:drivetrain} respectively). Besides modeling the physical architecture of the system-of-interest, the function behavior in logical terms (i.e. implementation independent manner) is captured too. The ``braking behavior'' is described below.

The \textit{Operator} sends the signals to \textit{Requester}, the \textit{`braking request'} is created. \textit{Requester} represents the braking pedal. It creates a ``propulsion torque event'', with return value of braking torque. The \textit{`braking request'} is then forwarded to \textit{Powertrain} which communicates with \textit{Wheels}, \textit{Engine} and \textit{Drivetrain}. \textit{Powertrain} forwards the \textit{`braking request'} to the \textit{Drivetrain}, where \textit{braking} happens in actual. It is consisted of three subsystems: \textit{Generator}, \textit{Dclink} and four \textit{Hub Units}. The \textit{Drivetrain} transform, transmit and modulate mechanical input torque to actualize requested mechanical torques at wheel interfaces. After receiving the
\textit{`braking request'}, engine requests how much power needs to be generated and sends that to \textit{Generator} through the \textit{Drivetrain}. \textit{Generator} transfers mechanical energy to electrical and forwards it to \textit{DcLink}, and then \textit{DcLink} transmits that energy and forwards it to four \textit{hub units} which modulate it and send to \textit{Wheels}.

The operations and signals identified to realize (logically) the ``braking behavior'' is represented in a block diagram as shown in Fig.~\ref{fig:drivetrain} (the color scheme to be explained later). It can be noted that the model elements (structure) illustrated are named with prefix ``Brake'' to indicate the modeling is focused only on ``analyzing'' and capturing the ``Braking'' behavior in isolation. The structure indicates the logical decomposition or ``allocation'' of the function w.r.to
the overall product breakdown structure. This is an important phase for both requirement and system engineers, and enhance communications within development teams. The results of this phase are also reviewed with the stakeholders.

\subsection{System Modeling (Behavior)}
It is the next level of modeling activity using SysML Statemachine diagrams to
capture the functional behavior of the structural ``parts'' (w.r.to the system-of-
interest). While in this paper, we restrict the behavior specification to that of
``Braking'', this specification is ``incrementally'' developed by considering each
of the machine functions separately and eventually ``combined'' (manually). For
instance, in Fig.~\ref{fig:statemachine}, the behavior of the HubUnit part of BrakeDrivetrain is presented. It can be noted the granularity of the (behavior) modeling effort is not arbitrary, but carefully chosen to cover the ``new'' parts, in this case the HubUnit (electrically steering the wheels).

As result of the modeling activity described in previous subsection, it can be
noted, there will be multiple state machines for the \textit{HubUnit} (in other words,
multiple \textit{HubUnits} each pre-fixed with the individual function names). These
state machines are ``combined'' to create a single state machine, at suitable
granularity, that serves as the System Specification (for the DLS). Thus, this
modeling phase also contributes to the overall system design decisions. For ex-
ample, as shown in Fig.~\ref{fig:statemachine}, the color scheme indicates the design decision to implement the corresponding operation in \textit{Software} or \textit{Hardware} (further explained in subsection below).

\subsection{Modeling System Design (S/W \& H/W Views)}
\begin{figure*}[t]
\centerline{\includegraphics[height=2.14in,width=5.2in,angle=0]{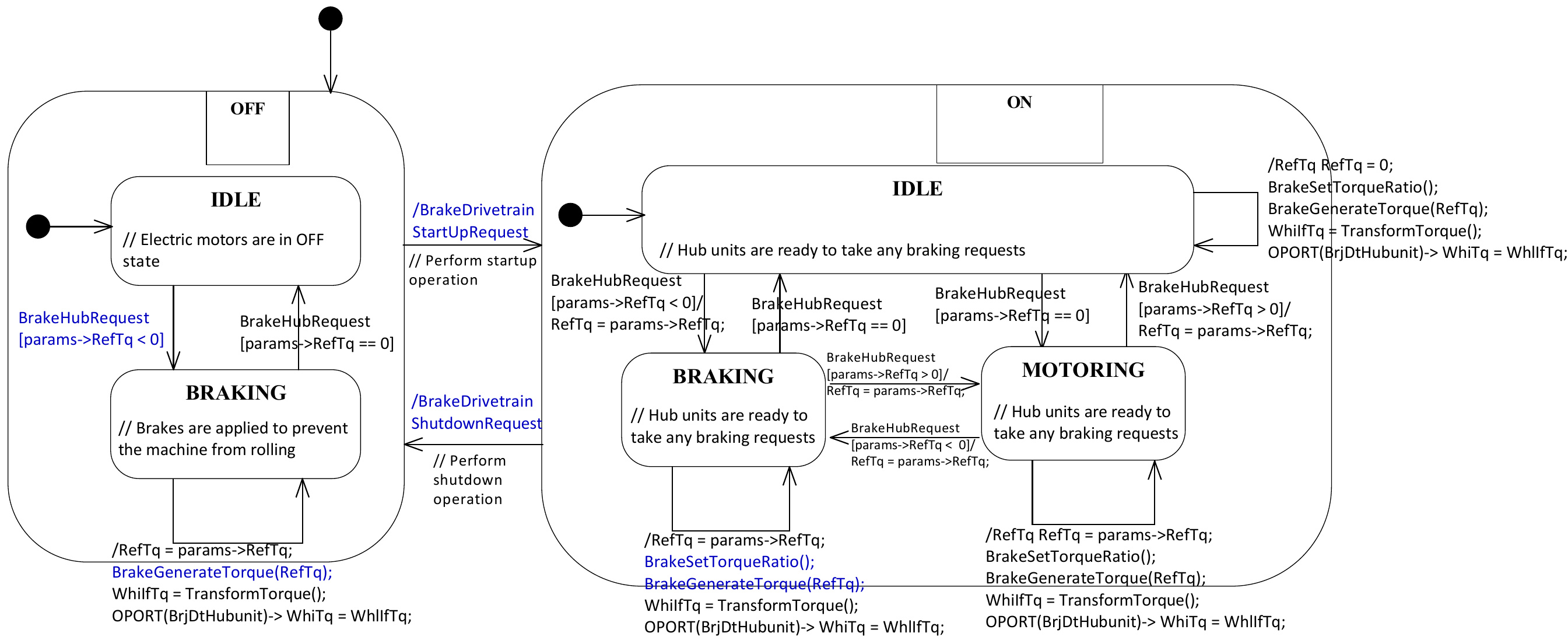}}
\vspace*{-1.6ex}
\caption{State Machine of \textit{BrakeHubUnit} subsystem}
\vspace*{-2.0ex}
\label{fig:statemachine}
\end{figure*}

This is the final phase of the SysML-modeling activity and concerns the detailed design modeling. As described in the previous subsection, the ``software'' and ``hardware'' parts are identified during behavior modeling phase, as part of the system specification. This phase further requires extensive domain expertise (both system and software level), for the technical trade-offs to be made regarding whether a certain SysML Operation is to be implemented in software or hardware (e.g. sensor). For instance, as shown in Fig.~\ref{fig:statemachine}, the \textit{blue} color represents the design decision that the corresponding operation has been allocated to the software modules and \textit{red} color operations allocated to the hardware components for realization of the overall system. Based on these, allocation views are modeled in SysML (skipped in this paper due to lack of space).

\section{Conclusions}
The SysML-based modeling approach presented in this paper is primarily a \textit{reverse-engineering} effort in capturing functionality from legacy implementations in the hardware and software. However, the approach is generic enough to
be extended into a useful modeling approach complimentary to existing software development approaches.

\section*{Acknowledgments}
This work is partially funded from the Electronic Component Systems for Euro-
pean Leadership Joint Undertaking under grant agreement No. 737494 and The
Swedish Innovation Agency, Vinnova (MegaM@Rt2).
 

%
%

\begin{thebibliography}{1}

\bibitem{friedenthal2007incose}
Friedenthal, S., Griego, R., Sampson, M.: INCOSE model based systems
  engineering (MBSE) initiative. In: INCOSE 2007 Symposium (2007)

\bibitem{friedenthal2014practical}
Friedenthal, S., Moore, A., Steiner, R.: A practical guide to SysML: the
  systems modeling language. Morgan Kaufmann (2014)

\end{thebibliography}

\end{document}